\documentclass[conference]{IEEEtran}
\IEEEoverridecommandlockouts
\usepackage{cite}
\usepackage{graphicx}
\usepackage{cite}
\usepackage{amsmath,amssymb,amsfonts}
\usepackage{algorithmic}
\usepackage{textcomp}
\usepackage{xcolor}
\usepackage{fancyhdr}
\usepackage{fancyvrb}
\usepackage[ruled,vlined]{algorithm2e}
\usepackage{amsthm}
\usepackage{enumitem}
\usepackage{tablefootnote}
\usepackage{multirow}
\usepackage{colortbl}
\definecolor{Gray}{gray}{0.65}
\begin{document}

\title{
HyGNN: Drug-Drug Interaction Prediction via Hypergraph Neural Network
}

\author{
\IEEEauthorblockN{Khaled Mohammed Saifuddin\textsuperscript{*}, Briana Bumgardner\textsuperscript{¶}, Farhan Tanvir\textsuperscript{§}, Esra Akbas\textsuperscript{†}}

\IEEEauthorblockA{\textit{\textsuperscript{*†}Georgia State University}, USA\\
\textit{\textsuperscript{¶} Rice University}, USA \\
\textit{\textsuperscript{§} Oklahoma State University}, USA \\
\textsuperscript{*}ksaifuddin1@student.gsu.edu,
\textsuperscript{¶}bb64.edu@rice.edu, 
\textsuperscript{§}farhan.tanvir@okstate.edu, 
\textsuperscript{†}eakbas1@gsu.edu
}
}

\maketitle
\begin{abstract}
Drug-Drug Interactions (DDIs) may hamper the functionalities of drugs, and in the worst scenario, they may lead to adverse drug reactions (ADRs). Predicting all DDIs is a challenging and critical problem. Most existing computational models integrate drug-centric information from different sources and leverage them as features in machine learning classifiers to predict DDIs. However, these models have a high chance of failure, especially for new drugs when all the information is not available. This paper proposes a novel \textit {Hypergraph Neural Network} (\texttt{HyGNN}) model based on only the Simplified Molecular Input Line Entry System (SMILES) string of drugs, available for any drug, for the DDI prediction problem. To capture the drug chemical structure similarities, we create a hypergraph from drugs' chemical substructures extracted from the SMILES strings. Then, we develop \texttt{HyGNN} consisting of a novel attention-based \textit{hypergraph edge encoder} to get the representation of drugs as hyperedges and a decoder to predict the interactions between drug pairs. Furthermore, we conduct extensive experiments to evaluate our model and compare it with several state-of-the-art methods. Experimental results demonstrate that our proposed \texttt{HyGNN} model effectively predicts DDIs and impressively outperforms the baselines with a maximum F1 score, ROC-AUC, and PR-AUC of 94.61\%, 98.69\%, and 98.68\%, respectively. Finally, we show that our models also work well for new drugs.     
\end{abstract}


\section{Introduction}

Many patients, especially those who suffer from chronic diseases such as high blood pressure, cancer, and heart failure, often consume multiple drugs concurrently for their disease treatment. Simultaneous usage of multiple drugs may result in Drug-Drug Interactions (DDIs). These interactions may unexpectedly reduce the efficacy of drugs and even may lead to adverse drug reactions (ADRs) \cite{Han2022,tan20224sdrug}. Therefore, it is important to identify potential DDIs early to minimize these adverse effects. 
However, since clinical trials to identify DDIs are performed on a few patients for a brief period \cite{Qiu2021}, many potential new drug DDIs remain undiscovered before it is open to the market. Also, it is too expensive to do clinical experiments with all possible drug pairs. Thus, there is an obvious need for a computational model to detect DDIs and mitigate unanticipated reactions automatically. 
 
With the availability of public databases, including drug-related information like DrugBank\footnote{https://go.drugbank.com/}, STITCH\footnote{http://stitch.embl.de/}, SIDER\footnote{http://sideeffects.embl.de/}, PubChem\footnote{https://pubchem.ncbi.nlm.nih.gov/}, KEGG\footnote{https://www.kegg.jp/}, etc., different computational models have been proposed to detect DDIs~\cite{yi2022graph,Zitnik2018}. Some of these models consider drug pairs' chemical structure SMILES similarity
or binding properties \cite{Takeda2017}. SMILES is a specification that uses ASCII characters to define molecular structures explicitly. With a string of characters, SMILES may depict a three-dimensional chemical structure.
On the other hand, out of the entire chemical structure, only a few substructures are responsible for chemical reactions between drugs, and the rest are less relevant~\cite{Huanga}. However, considering the whole chemical structure may create a bias toward irreverent substructures and thus undermine the DDIs prediction ~\cite{Zheng2019}. With the increasing ability of more relational information about drugs, most of the current methods integrate multiple data sources to extract drug features such as side effects, target protein, pathways, and indications~\cite{Tanvir2021,qiu2021comprehensive}. 

Network-based methods have recently been explored
in this domain, where drug networks are constructed based on drugs' known DDIs. Most of the graph-based methods consider a dyadic relationship between drugs. They operate on a simple regular graph where each vertex is a drug, and each edge shows a connection between two nodes. However, some methods also consider the relations of drugs to other biological entities to create heterogeneous graphs. Then, different topological information is extracted from the network to predict unknown links (i.e., interactions) between drugs. 
With the current advancement of the graph neural network (GNN), different GNN models for DDI prediction problems are proposed~\cite{ Zitnik2018, Feng2020}. While some of them create heterogeneous graphs manually from different resources, some of them create biomedical knowledge graphs by extracting triples from raw data (e.g., DrugBank)~\cite{ren2022biodkg, yu2021sumgnn, Yu2021}. In these graphs, different entities, such as drugs, proteins, and side effects, are represented as nodes, and relations between those entities are represented as the edges.
While including multiple drug-centric information from different sources would help to learn about DDI and have achieved strong performance, it is challenging to integrate data from different resources. It is also tough to interpret which information is the most valuable in DDI prediction which needs a strong knowledge of biomedical entities and this is challenging for drugs in the early development stage. Moreover, multiple information may not always be available for all drugs, specially new drugs; therefore, these models may fail whenever any information is unavailable~\cite{Mei2021}.

In this paper, to address these problems, we present a novel GNN-based approach for DDI prediction by only considering the SMILES string of the drugs, which is available for all drugs. 
Our method relies on the hypothesis that similar drugs behave similarly, are likely to interact with the same drugs, and two drugs are similar if they have similar substructures as functional groups in their SMILES strings \cite{vilar2012drug, martin2002structurally}. Finding similarities between SMILES strings based on their common substructures is a challenging task.
To properly depict the structural-based similarity between drugs, we present them in a hypergraph setting, representing drugs as hyperedges connecting many substructures as nodes. A hypergraph is a unique model of a graph with hyperedges. Unlike a regular graph where the degree of each edge is 2, hyperedge is degree-free; it can connect an arbitrary number of nodes~\cite{bretto2013hypergraph,Feng2018,aktas2021hypergraph, aktas2021identifying}. After constructing the hypergraph, we develop a Hypergraph Neural Network (\texttt{HyGNN}), a model that learns the DDIs by generating and using the representation of hyperedges as
drugs. \texttt{HyGNN} has an encoder-decoder architecture. First, we present a novel \textit{hypergraph edge encoder} to generate the embedding of drugs. Afterward, the pair-wise representations of drugs are passed through decoder functions to predict a binary score for each drug pair that represents whether two drugs interact. The main contributions of this paper are summarized as follows:
\begin{itemize}
\item \textbf{Hypergraph construction from SMILES strings: } 
We construct a novel hypergraph to depict the drugs' similarities. In the hypergraph, while each substructure extracted from the drugs' SMILES strings is represented as a node, each drug, consisting of a set of unique substructures, is represented as a hyperedge. This hypergraph represents the higher-level connections of substructures and drugs, which helps us to define complex similarities between chemical structures and drugs. Also, this helps us to learn better representation for drugs with GNN models with a passing message not only between 2 nodes but between many nodes and also between nodes and hyperedges.
\item \textbf{Hypergraph GNN:}
To learn and predict DDIs, we propose a novel hypergraph GNN model, called \texttt{HyGNN}, consisting of a novel hyperedge encoder and a decoder. Encoder exploits two layers where the first layer generates the embedding of nodes by aggregating the embedding of hyperedges. Then, the second layer generates the embedding of hyperedges (i.e., drugs) by aggregating the embedding of nodes. Since not all but a few substructures are mainly significant in chemical reactions, we use attention mechanisms to learn the significance of substructures (nodes) for drugs (edge) and chemical reactions.
Furthermore, a decoder is modeled to predict the DDIs by taking the pair-wise drug representations as input. 
 Our method solely utilizes drugs' chemical structure data to predict DDIs without requiring any other information or strong knowledge of biomedical entities. Chemical structures are obtained from SMILES strings that any drug has. Therefore, our model is applicable to any drugs, including new drugs, without other information, such as side effects and DDIs. 
\item \textbf{Extensive experiments: } We conduct extensive experiments to compare our proposed model with the state-of-the-art models. The results with different accuracy measures show that our method significantly outperforms all the baseline models. Also, we show with case studies that our model can find not only new DDIs for existing drugs but also DDIs for new drugs.
\end{itemize}
The rest of the paper is structured as follows. First, we briefly review the related work on DDI and hypergraph GNN in Section~\ref{sec:works}. Section~\ref{sec:method} describes our proposed \texttt{HyGNN} model, including hypergraph construction from SMILES strings, encoder and decoder layers of the \texttt{HyGNN} model works for generating the drug embedding and predicting DDIs. Next, in Section~\ref{sec:experiment}, we describe the experimental results and discussion. Finally, a conclusion is given in Section~\ref{sec:con}.

\section{Related Work}
\label{sec:works}
Many works have been proposed for the DDI
prediction problem over the years. It can be categorized into similarity-based, classification-based, and network-based methods. Previous works assume that similar
drugs have similar interaction profiles and define different similarities between drugs. Traditionally, pharmacological, topological, or semantic similarity based on statistical learning is utilized to predict DDIs. Vilar et al. \cite{Vilar2012} identify the DDIs based on molecular similarities. They represent each drug by
a molecular fingerprint, a bit vector reflecting
the presence/absence of a molecular feature. \cite {Gottlieb2012} develops INDI that uses seven different drug-drug similarity measures learned from drug side-effect, fingerprints, therapeutic effects, etc. Another vital research is \cite{Ferdousi2017} that incorporates four different biological information (e.g., target, transporter, enzyme, and carrier) of drugs to measure the similarity of drug pairs. 

Some models extract features of drugs from various biological entities and drug interaction information and apply different machine learning (ML) methods for DDI training~\cite{Ibrahim2021,Zheng2019, Davazdahemami2018ACP, Luo2014DDICPIAS}. Davazdahemami and Delen \cite{Davazdahemami2018ACP} constructed a graph containing both drug-protein and drug-side effect interactions. They also employed a classification method on the feature set and produced many similarities and centrality metrics based on the network. These features were then fed into four machine models. 
Luo et al. \cite{Luo2014DDICPIAS} propose a DDI prediction server that provides real-time DDI predictions based only on molecular structure. The server docks a drug's chemical structure across 611 human proteins to create a 611-dimensional docking vector. The drug pair features are created by concatenating the docking vectors for drug pairings. Finally, utilizing these features, a logistic regression model for DDI prediction is developed. Ibrahim et al.\cite{Ibrahim2021} first extract different similarity features and employ logistic regression to pick the best feature; later, the best feature is used in six different ML classifiers to predict DDIs. One primary problem in DDI prediction is the lack of negative samples. A solution to address this problem is proposed in \cite{Zheng2019}, named DDI-PULearn. It first generates negative samples using one-class SVM and kNN. Then positive and negative samples are used to predict DDIs.

Last decade, network-based models got great attention for drug-related problems. Some researchers construct a drug network using known DDIs where drugs are nodes, and interacting drugs are connected by a link \cite{Kastrin2018}. Moreover, heterogeneous information networks leveraging different biomedical entities, such as proteins, side effects, pathways, etc., are also used to address similar problems \cite{Tanvir2021}. As a different model, \cite{Chen2019} constructs a molecular graph for each drug from its SMILES representation. Moreover, existing network-based models often extract drug embedding and directly learn latent node embedding using various embedding methodologies. As a result, their capacity to obtain specific neighborhood information on any organization in KG is restricted. 

Recently, GNN has shown promising performance in different fields that include drug discovery \cite{Gaudelet2021}, drug abuse detection \cite{Saifuddin2022}, and drug-drug interaction \cite{Bai2020, Bumgardner2022, Chen2019}, etc. Decagon \cite{Zitnik2018} created a knowledge graph based on protein-protein, drug-drug, and drug-protein interactions. They also created a relational graph convolutional neural network for predicting multi-relational links in multimodal networks. Furthermore, they used a novel graph auto-encoder technique to create an end-to-end trainable model for link prediction on a multimodal graph. CASTER \cite{Huanga} recently created
a dictionary learning framework for predicting DDIs based on drug chemical structures. They predict drug-drug interactions using the drug's molecular structure in a text format of SMILES \cite{Weininger1988SMILESAC} strings representation and outperform numerous deep learning approaches such as DeepDDI \cite{Ryu2018DeepLI} and molVAE \cite{GmezBombarelli2018AutomaticCD}. CASTER uses a sequential pattern mining approach to identify the common substrings included in the SMILES strings supplied during the training phase, which are then translated to an embedding using an encoder module. These features are then transformed into linear coefficients fed to a decoder and a predictor to
yield DDI predictions. \cite{Bumgardner2022} constructs a drug network where two drugs are connected if they share common chemical substructures. Then, they apply different GNN models on the network to get the representation of drugs, and drug pair representation is passed to the ML classifier to predict interaction. 

More recently, hypergraph and hypergraph neural network models have been developed to capture higher-order relations between different objects~\cite{Ding2020, Feng2018, Chen2020a, Hwang2021}. All these works learn the representation of nodes and use these for node classification problems. Our \texttt{HyGNN} model is the first attempt to use hypergraph structure for DDI problems and, in general, drug-related problems. Also, our model aims to learn the representation of hyperedges and edge pair classification, which is different from current hypergraph neural network models.

\section {\texttt{HyGNN} model for DDI}
\label{sec:method}

In this section, we first define our DDI prediction problem and then summarize the preliminaries, model, and settings (Section \ref{sec:PF}). Then we explain our hypergraph construction step with substructures extraction from Drugs (Section \ref{sec:HGC}). After that, we introduce our proposed \texttt{HyGNN} model with attention-based encoder and decoder layers (Section \ref{sec:HYGNN}).

\subsection{Problem Formulation}
\label{sec:PF}

Our goal is to develop a computational model that takes a drug pair $(D_{x}, D_{y})$ as input and predicts whether there exists an interaction between this drug pair. Each drug is represented by the SMILES string. SMILES is a unique chemical representation of a drug that consists of a sequence of symbols of molecules and the bonds between them. 

Most of the graph-based existing DDI methods consider a dyadic relationship between drugs. This simple graph type considers an edge that can connect a maximum of two objects. However, there could be a more complex network in real life where an arbitrary number of nodes may interact as a group, so they could be connected through a hyperedge (i.e., triadic, tetradic, etc.). A hypergraph can be used to formulate such a complex network. A formal definition of the hypergraph is given below.

\emph{\textbf{Hypergraph:}}
\emph{A hypergraph is a special kind of graph defined as $G = (V, E)$ where $V =\{v_{1},....,v_{m}\}$ is the set of nodes and $E = \{e_{1}.....e_{n}\}$ is the set of hyperedges. Each hyperedge $e_{j}$ is degree-free and consists of an arbitrary number of nodes.}

Like the adjacency matrix of a regular graph, a hypergraph can be denoted by an incidence matrix $H$ with $H_{i,j}=1$ if the node $i$ is in the hyperedge $j$ as $v_{i}\in e_{j}$ and $H_{i,j}=0$ otherwise. 
\begin{figure*}[t!]
    \centering
  \includegraphics[width=15cm]{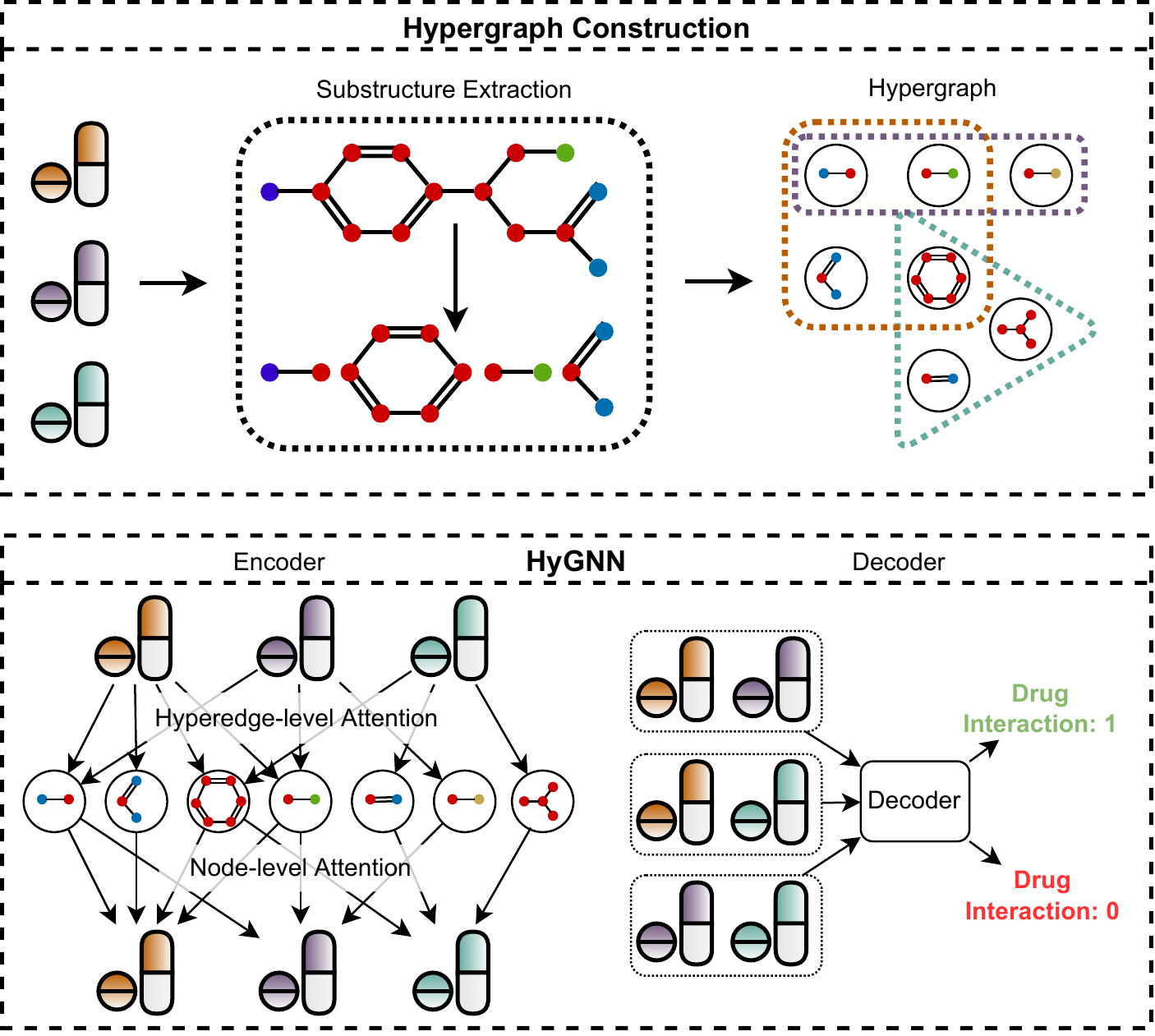}
   \caption{System architecture of the proposed method. The First step is to construct a hypergraph network of drugs where each drug is a hyperedge, and frequent chemical substructures of drugs are the nodes. The second step is to design a hypergraph neural network (\texttt{HyGNN}) model with an attention-based encoder for hyperedge (drug) representation learning  and decoder for DDI learning}
    \label{fig:method}
\end{figure*}

In this paper, we construct a hypergraph network of drugs where each drug is a hyperedge, and the chemical substructures of drugs are the nodes. The chemical structures of a drug can be obtained from the SMILES, a unique chemical representation of a drug. We design a novel hypergraph neural networks
model as an encoder-decoder architecture to accomplish the DDI prediction task. The encoder part exploits an attention mechanism to learn the representations of hyperedges (drugs) by giving attention to edges and nodes (substructures). The decoder predicts the interaction between drug pairs using latent learned drug features. The whole system is trained in a semi-supervised fashion. The functional architecture of this paper is shown in Figure 1.
Our proposed model consists of two steps:
\begin{enumerate}
    \item Hypergraph construction from SMILES,
    \item DDI prediction with hypergraph neural networks
\begin{enumerate}
 \item Encoder: Drug (Hyperedge) representation learning  
    \item Decoder: DDI prediction
\end{enumerate}\end{enumerate}

\subsection{Drug Hypergraph Construction}\label{sec:HGC}

We construct a hypergraph to depict the structural similarities among drugs. At first, we decompose all the drugs' SMILES into a set of substructures. In our drug hypergraph, these substructures are used as nodes. Moreover, each drug with a certain number of substructures is represented by a hyperedge. Drugs as hyperedges may connect with other drugs employing shared substructures as nodes. This hypergraph represents the higher-level connections of substructures and drugs, which may help to define complex similarities between chemical structures and drugs. Also, this helps us to learn better representation for drugs with GNN models with the passing message not only between 2 nodes but between many nodes and also between nodes and edges. 
 
Substructures can be generated by utilizing different algorithms such as ESPF \cite{Huangb}, $k$-mer\cite{Zhang2017}, strobemers \cite{Sahlin2021}, etc. In this project, we use ESPF and $k$-mer to see the effect of substructures on the results. While $k$-mer use all extracted substructures, ESPF selects the most frequent ones. Algorithm \ref{algorithm:1} briefly shows the hypergraph construction steps.

\SetKwComment{Comment}{/* }{ */}
\begin{algorithm}
\SetAlgoLined
\textbf{Input: } $SMILES\_strings$\\
 \textbf{Output: }Hypergraph incident matrix: $H$\\

Call $Substructure\_Decom(SMILES\_strings)$\Comment*[r]{Substructure\_Decom() could be ESPF or $k$-mer that decomposes SMILES into substructures. It returns a list of unique substructures and drug dictionary that will be used in the following for loop}
 
 \For {each Substructure in Substructure\_list}
 {
      \If {Substructure is in Drug\_dict[SMILES]}{
      $H[i,j]=1$ \Comment*[r]{i,j is the id of substructure and drug, respectively.} }
 }
\textbf{Output: }$H$, Hypergraph incident matrix
\caption{Drug Hypergraph Construction }
\label{algorithm:1}
\end{algorithm}

\SetKwComment{Comment}{/* }{ */}
\begin{algorithm}[h]
\SetAlgoLined
\textbf{Input: } Set of initial SMILES tokens ${S}$ as atoms and bonds, set of tokenized SMILES strings ${TS}$, frequency threshold $\alpha$, and size threshold ${L}$ for $S$.\\
 \For {${t = 1\dots,L}$}{
  $(S1, S2)$, $f$ $\leftarrow$ scan $TS$ \Comment*[r]{$(S1, S2)$ is the frequentest consecutive tokens.}
  \If{$f < \alpha$}
  {
   break \Comment*[r]{$(S1, S2)$'s frequency lower than threshold}
   }
   $TS \leftarrow$ find $(S1, S2) \in $ $TS$, replace with $(S1S2)$ \Comment*[r]{update $TS$ with the combined token $(S1S2)$}
   $S \leftarrow S \cup (S1S2)$ \Comment*[r]{add $(S1S2)$ to the token vocabulary set $S$}
 }
\textbf{Output: } $TS$, the updated tokenized drugs; $S$, the updated token vocabulary set.
\caption{Explainable Substructure Partition Fingerprint (ESPF)}
\label{algorithm:2}
\end{algorithm}

 \textbf{ESPF:} Like the concept of sub-word units in the natural language processing domain, ESPF is a powerful tool that decomposes sequential structures into interpretable functional groups. They consider that a few substructures are mainly responsible for drug chemical reactions, so they
extract frequent substructures as influential ones. The ESPF algorithm is shown in Algorithm \ref{algorithm:2}.
 Given a set of drug SMILES strings $S$, ESPF finds the frequent repetitive moderate-sized substructures from the set and replaces the original sequence with the substructures. If the frequency of each substructure in $S$ is above a predefined threshold, it is added to a substructure list as a vocabulary. 
Substructures appear in this list in the most to least frequent order. We use this vocabulary list as our nodes and decompose any drug into a sequence of frequent substructures concerning those. For any given drug, we partition its SMILES in order of frequency, starting from the highest frequency.
An example of partitioning a SMILES of a drug, DB00226, is as follows.
\begin{center}
\texttt{ NC(N)=NCC1COC2(CCCCC2)O1 }

$\Downarrow$

\texttt{ \underline{N} \underline{C(N)} \underline{=N} \underline{CC1} \underline{CO} \underline{C2} \underline{(CCC} \underline{CC2)} \underline{O1}.}
\end{center}

\textbf{$\textbf{k}$-mer:} $k$-mer is a tool to decompose sequential structures into subsequences of length $k$. It is widely used in biological sequence analysis and computational genomics. Similar to n-gram in natural language processing, a k-mer is a sequence of k characters in a string (or nucleotides in a DNA sequence). To get all k-mers from a sequence, we need to get the first k characters, then move just a single character to start the next k-mer, and so on. Effectively, this will create sequences that overlap in k-1 positions. Pseudocode for $k$-mer is shown in Algorithm \ref{algorithm:3}.

For a sequence of length $l$, there are $l-k+1$ numbers of $k$-mers and $n^k$ total possible number of $k$-mers, where $n$ is the number of monomers. $k$-mers are like words of a sentence. $k$-mers help to bring out semantic features from a sequence. For example, for a sequence \texttt{NCCO},
 monomers: \{N, C, and O\}, 2-mers: \{NC, CC, CO\} , 3-mers: \{NCC, CCO\}.

\SetKwComment{Comment}{/* }{ */}
\begin{algorithm}[t]
\SetAlgoLined
\textbf{Input: } $SMILES\_strings$, size threshold $k$\\
Substructure\_list: []\\
Drug\_dict:\{\}\\
\For {each $SMILES$ in $SMILES\_strings$}{
Lst=[]\\
\For {$x$ in range ($l$-$k$+1)} 
{\tcc{$l$ is the length of $SMILES$}
$C=SMILES[x:x+k]$\\
$Lst.append(C)$\\
$Substructure\_list.append(C)$
}
$Drug\_dict[SMILES]=Lst$
}
\textbf{Output: } Drug\_dict, Substructure\_list
\caption{$k$-mer}
\label{algorithm:3}
\end{algorithm}

\subsection{Hypergraph Neural Network (\texttt{HyGNN}) for DDI Prediction}\label{sec:HYGNN}

We utilize the Hypergraph Neural Network (\texttt{HyGNN}) for DDI prediction. \texttt{HyGNN} includes an encoder, which generates the embedding of drugs, and a decoder that uses the embedding of drugs from the encoder to predict whether a drug pair interact or not.

\subsubsection{\textbf{Drug Representation learning via \texttt{HyGNN} - Encoder}}
To detect interacting pairs of drugs, we need features of drug pairs and, thus, features of drugs that encode their structure information. To generate features of drugs, we propose a novel \textit{hypergraph edge encoder}. It creates ${d}^{'}$ dimensional embedding vectors for hyperedges (drugs) instead of nodes as in regular GNN models. Given the edge feature matrix, $F \in R^{E*d}$ and incidence matrix $H \in R^{V*E}$, the encoder of the \texttt{HyGNN} generates a feature vector of $d'$ dimension through learning a function $Z$. Any layer (e.g., ${(l+1)^{th}}$ layer) of \texttt{HyGNN} can be expressed as
\begin{equation}
F^{l+1} = Z(F^{l},H^T).
\end {equation}    

We consider the \textit{hypergraph edge encoder} with a memory-efficient self-attention mechanism. It consists of two different levels of attention:
(1) hyperedge-level attention, (2) node-level attention.

While hyperedge-level attention aggregates the hyperedge information to get the representation of nodes, node-level attention layer aggregates the connected vertex information to get the representation of hyperedges. In general, we define the \texttt{HyGNN} attention layers as \\
\begin{align}
p_{i}^{l}=\mathbf{A_{E}}^{l}(p_{i}^{l-1},{q_{j}^{l-1}}|\forall e_{j}\in E_{i}),\\
q_{j}^{l}=\mathbf{A_{V}}^{l}(q_{j}^{l-1},{p_{i}^{l}}|\forall v_{i}\in e_{j})
\end{align}
where \textbf{A\textsubscript{E}} is an edge aggregator that aggregates features of hyperedges to get the representation $p_{i}^l$ of node $v_i$ in layer-$l$ and $E_{i}$ is the set of hyperedges that are connected to node $v_i$. Similarly, \textbf{A\textsubscript{V}} is a node aggregator that aggregates features of nodes to get the representation $q_{j}^l$ of hyperedge $e_j$ in layer-$l$ and $v_{i}$ is the node that connects to hyperedge $e_j$.

\textit{Hyperedge-level attention:} In a hypergraph, each node may belong to multiple numbers of edges. However, the contribution of hyperedges to a node may not be equal. That is why we design an attention mechanism to highlight the crucial hyperedges and aggregate their features to compute the node feature $p_{i}^l$ of node $v_i$. With the attention mechanism, $p_{i}^l$ is defined as
 
\begin{equation}
\label{eqn_n_4}
p_{i}^{l}=\alpha \left(\sum_{e_{j}\in {E_{i}}}Y_{ij}W_{1}q_{j}^{l-1}\right)
\end{equation}
where $\alpha$ is a nonlinear activation function, $W_1$ is a trainable weight matrix that linearly transforms the input hyperedge feature into a high-level, $E_i$ is the set of hyperedges connected to node $v_i$, and $Y_{ij}$ is the attention coefficient of hyperedge $e_j$ on node $v_{i}$.
The attention coefficient is defined as

\begin{equation}
    Y_{ij}= \frac{\exp({\mathbf{e_j}})}{\sum_{e_k\in E_i}\exp(\mathbf{e_k})}
\end{equation}

\begin{equation}
    \mathbf {e_{j}} = \beta {(W_{2}q_j^{l-1}*W_{3}p_i^{l-1})}
\end{equation}
where $\beta$ is a LeakyReLU activation function, $W_2$, $W_3$ are the trainable weight matrices, and $*$ is the element-wise multiplication.

\textit{Node-level attention:} Each hyperedge in a hypergraph consists of an arbitrary number of nodes. However, the importance of nodes in a hyperedge construction may not be the same. We design a node-level attention mechanism to highlight a hyperedge's important nodes and aggregate their features accordingly to compute the hyperedge feature $q_{j}^l$ of hyperedge $e_j$. With the attention mechanism, $q_{j}^l$ is defined as
\begin{equation}
\label{eqn_n_7}
q_{j}^{l}=\alpha \left(\sum_{v_{i}\in {e_{j}}}X_{ji}W_{4}p_{i}^{l}\right)
\end{equation}
where $W_4$ is a trainable weight matrix, and $X_{ji}$ is the attention coefficient of node $v_i$ in the hyperedge $e_{j}$.
The attention coefficient is defined as

\begin{equation}
    X_{ji}= \frac{\exp({\mathbf{v_i}})}{\sum_{v_k\in e_j}\exp( \mathbf{v_k})}
\end{equation}

\begin{equation}
    \mathbf {v_{i}} = \beta {(W_{5}p_i^{l}*W_{6}q_{j}^{l-1})}
\end{equation}
where $v_{k}$ is the node that belongs to hyperedge $e_{j}$, $W_5$, $W_6$ are the trainable weight matrices.

Our hypergraph edge encoder model works based on these two attention layers that can capture high-order relations among data. Given the input hyperedge features, we first gather them to get the representation of nodes with hyperedge-level attention, then we gather the obtained node features to get the representation of hyperedges with node-level attention.

\subsubsection{\textbf{DDI prediction - Decoder}}
 After getting the representation of drugs from the encoder layer, our target is to predict whether a given drug pair interacts or not. To accomplish this target, we design a decoder. 

Given the vector representations $(q_{x},q_{y})$ of drug pairs $(D_{x},D_{y})$ as input, the decoder assigns a score, $p_{x,y}$ to each pair through a decoder function defined as \begin{equation}p_{x,y}=\gamma(q_{x},q_{y})\end{equation} We use two different types of decoder functions:

\textbf{MLP:} After concatenating the features of drug pairs, we pass it through a multi-layer perceptron (MLP), which returns a scalar score for each pair
\begin{align}
\begin{aligned}
\gamma(q_x,q_y) = f_2(f_1(q_x \mathbin\Vert q_y))\\
\end{aligned}
\end{align}
where $f_1$ and $f_2$ are two different layers of MLP, and $||$ is the concatenation operation.

\textbf{Dot product:} We compute a scalar score for each edge by performing element-wise dot product between features of drug pairs using
\begin{equation}
\gamma(q_x,q_y) = q_x \cdot q_y
\end{equation}

Afterward, we pass the decoder output through a sigmoid function $\sigma({\gamma(q_x,q_y)})$
that generates predicted labels, $Y'$, within the range 0 to 1. Any output value closer to 1 implies
a high chance of interaction between two drugs. 

\subsubsection {\textbf{Training the whole model}} We consider the DDI prediction problem as a binary classification problem predicting whether there is an interaction between drug pairs or not. As a binary classification problem, we train our entire encoder-decoder architecture using a binary cross-entropy loss function defined as
\begin{equation}
loss= -\sum_{i=1}^{N}\left({Y_{i}\log Y_{i}^{'}+(1-Y_{i})\log(1-Y_{i}^{'})}\right)
\end{equation}
where $N$ is the total number of samples, $Y_{i}$ is the actual label, and $Y_{i}^{'}$ is the predicted label.

\subsection{Complexity}
Our method is highly efficient with paralyzing across the edges and nodes~\cite{velivckovic2017graph}. The hyperedge encoder generates ${d}^{'}$ dimensional embedding vector for each hyperedge with a given initial feature as $d$ dimensional vector using two-level attention. Hence, the time complexity in the encoder part is the cumulative complexity of attention layers. According to equation \ref{eqn_n_4}, the complexity in the hyperedge-level attention can be expressed as: $O(|E|d{d}^{'}+|V|D{d}^{'})$, where $D$ is the average degree of nodes. Similarly, the complexity in the node-level attention can be expressed as:
$O(|V|d{d}^{'}+|E|B{d}^{'})$, where $B$ is the average degree of hyperedges.

\section{Experiment}
\label{sec:experiment}

In this section, we evaluate our proposed \texttt{HyGNN} model for DDI prediction with extensive experiments on two different datasets. We use F1, ROC-AUC, and PR-AUC accuracy metrics to compare our model's performances with the state-of-art baseline models. Making DDI
predictions for new drugs could be more challenging than existing drugs.
Therefore, we assess our model’s performance for both new and existing drugs as well. First, we describe our datasets, TWOSIDES and DrugBank, then we explain our experiments, and present and analyze our results.
\subsection{Dataset}

We evaluate the proposed model using two different sizes of datasets. One is a small dataset, and another one is a large dataset. (1) \textbf{TWOSIDES} is our small dataset. \textbf{TWOSIDES} was created using data from adverse event reporting systems. Common adverse effects, such as hypotension and nausea, occur in more than a third of medication combinations, but others, such as amnesia and muscular spasms, occur in only a few. 
We extract 645 approved drugs' information from \textbf{TWOSIDES}. Each drug is linked to its chemical structure (SMILES). There are 63,473 DDI positive labels for the selected drugs. (2) \textbf{DrugBank} is our large dataset and it is the largest dataset for drugs that is publicly available. It is a drug knowledge database that includes clinical information about drugs, such as side effects and (DDIs). \textbf{DrugBank} also includes molecular data, such as the drug’s chemical structure, target protein, and so on. From \textbf{DrugBank}, we retrieve information on 1706 approved drugs along with their SMILES strings and 191,402 DDI information. 
Both datasets are publicly available on Therapeutics Data Commons (TDC) \footnote{https://tdcommons.ai/}. The first unified platform, TDC, was launched to comprehensively access and assess machine learning across the entire therapeutic spectrum.

\begin{table}
    \centering \normalsize
      \caption{Statistics of Dataset}
      \begin{tabular}{|c | c |c |}
  \hline
 \cellcolor{gray!60}  \textbf{Dataset}& \cellcolor{gray!60}  \textbf{\# of Drug} & \cellcolor{gray!60}  \textbf{\# of DDI}\\\hline \hline 
       TWOSIDES & 645  & 63473
        \\         
 \hline 
        DrugBank & 1706&191402
        \\    \hline   
     
        \end{tabular}
        \label{tab1}
     \end{table}
All known DDIs in both datasets are our positive samples. However, to train our model, we need negative samples as well. Therefore, we randomly sample a drug pair from the complement set of positive samples for each positive sample. Thus, we ensure a balanced dataset of equally positive and negative samples for an individual dataset.

We apply the ESPF algorithm and $k$-mer separately to extract the substructures from the SMILES string of drugs. For ESPF, we notice that when a lower frequency threshold is set, it generates many substructures, some of which may be unimportant. However, when a more significant threshold value is set, it generates fewer substructures and may lose some critical substructures. These substructures are used as nodes in the hypergraph. To examine the impact of the frequency threshold and thus the number of nodes in the hypergraph learning, we choose five different threshold values from 5 to 25. 
For $k$-mer, we notice that typically with the increment of $k$, the number of substructures (i.e., nodes) also increases. Similarly, to examine the impact of the $k$ and thus the number of nodes in the hypergraph learning, we choose five different values of $k$ from 3 to 15. A statistic of both datasets is shown in Table \ref{tab1}. The number of nodes for different threshold values of ESPF and $k$-mer is given in Table~\ref{tab2} and Table~\ref{tab3} for each dataset. 

\subsection{Parameter Settings}
Each dataset is randomly split into three parts:  train (80\%), validation (10\%), and test (10\%). We repeat this five times and report the average performances in terms of F1-score, ROC-AUC, and PR-AUC. The optimal hyper-parameters are obtained by grid search based on the validation set. The ranges of grid search are shown in Table \ref{tab4}. 

We employ a single-layer \texttt{HyGNN} having two levels of attention. We use a LeakyReLU activation function in the encoder side and a ReLU activation function in the MLP predictor of the decoder side. 
During training, we simultaneously optimize the encoder and decoder using adam optimizer. Each model is trained for 2000 epochs with an early stop if there is no change in validation loss for 200 consecutive epochs. 

For the baselines in subsection: \ref{sec:baselines}, each GNN model is used as a two-layer architecture. All other parameters of each GNN are set by following their sources. For DeepWalk and node2vec, the walk length, number of walks, and window size are set to 100, 10, and 5, respectively. We use Logistic Regression as a simple ML classifier.

\begin{table}
    \centering
      \caption{\# of Nodes (N) in the Hypergraph based on parameters of the methods, ESPF and $k$-mer, for TWOSIDES Dataset}
      \normalsize
      \begin{tabular}{|c|c||c|c|} \hline
       \cellcolor{gray!60}\textbf{ESPF} & \cellcolor{gray!60}$|N|$ &  \cellcolor{gray!60}\textbf{$k$-mer} & \cellcolor{gray!60}$|N|$ \\ \hline\hline
        5  & 555 & 3  & 822
        \\   \hline    
         10 & 324 & 6 & 7025
        \\    \hline     
         15 & 249 & 9 & 14002
        \\    \hline   
         20 & 208 & 12 & 17351
        \\    \hline   
         25 & 187 & 15 & 18155
        \\     
        \hline
        \end{tabular}
        \label{tab2}
\end{table}

\begin{table}
    \centering
      \caption{\# of Nodes (N) in the Hypergraph based on parameters of the methods, ESPF, and $k$-mer, for DrugBank Dataset}
      \normalsize
      \begin{tabular}{|c|c||c|c|} \hline
       \cellcolor{gray!60}\textbf{ESPF} & \cellcolor{gray!60}$|N|$ &
       \cellcolor{gray!60}\textbf{$k$-mer} & \cellcolor{gray!60}$|N|$ \\
       \hline\hline
         5  & 1266 & 3  & 1296
        \\     \hline    
         10 & 729 & 6 & 11849 
        \\     \hline    
         15 & 550 & 9 & 29443
        \\     \hline  
         20 & 462 & 12 & 43634
        \\       \hline
         25 & 400 & 15 & 51315
        \\      
        \hline
        \end{tabular}
        \label{tab3}
\end{table}

\begin{table}
    \centering \normalsize
      \caption{Hyper-parameter Settings}
      \begin{tabular}{|c|c|}        
      \hline
       \cellcolor{gray!60} \textbf{Parameter} &  \cellcolor{gray!60} \textbf{Values}\\
      \hline \hline
        Learning rate  & 1e-2, 5e-2, 1e-3, 5e-3
        \\         \hline
        Hidden units & 32, 64, 128
        \\       \hline
        Dropout & 0.1, 0.5
        \\       \hline
        Weight decay & 1e-2, 1e-3
        \\          \hline
        \end{tabular}
        \label{tab4}
     \end{table}

\subsection{Baselines}\label{sec:baselines}

We compare our model performance with different types of state-of-the-art models. We categorize the baseline models into five groups based on the data representation and methodology.

\textit{1. Random walk-based embedding (RWE) on DDI graph } We construct a regular graph based on the drug interaction information called a DDI graph. Drugs are represented as nodes, and two drugs share an edge if they interact. After constructing the graph, we apply the random walk-based graph embedding methods to get the representations of drugs.
DeepWalk \cite{perozzi2014deepwalk} and Node2vec \cite{grover2016node2vec} are two well-known graph embedding methods. They are both based on a similar mechanism of `walk' on the graph traversing from one node to another. We apply DeepWalk and Node2Vec on the DDI graph and generate the embedding of nodes. Afterward, we concatenate drug embeddings to get the drug pair features and feed that into a machine learning classifier for binary classification.\\

\textit{2. GNN on DDI graph:} 
 After constructing the DDI graph as explained above, we apply three different GNN models with unsupervised settings; graph convolution network (GCN) \cite{kipf2016semi}, graph attention network (GAT) \cite{velivckovic2017graph}, and GraphSAGE \cite{hamilton2017inductive} to get the representations of drugs. These GNN models are obtained from DGL\footnote{https://docs.dgl.ai/}. After getting the representations of drugs, we concatenate them pair-wise and use them as the features of drug pairs in the ML classifier for binary classification.
 \\

\textit{3. GNN on substructure similarity graph (SSG) }
We follow \cite{Bumgardner2022} to create the substructure similarity graph (SSG). We construct an edge between two drugs if they have at least a predefined number of common substructures. We apply the ESPF algorithm to the SMILES strings of drugs to get the frequent substructures. Afterward, we apply three different GNN models, GCN, GAT, and GraphSAGE, to the constructed graph to get the representations of drugs. The drug representations are then concatenated pair-wise and fed into a classifier to predict the DDI.
\\

\textit{4. CASTER} We apply the Caster algorithm~\cite{Huanga} for DDI prediction. It takes SMILES strings as input and employs frequent sequential pattern mining to discover the recurring substructures. They use the ESPF algorithm to extract frequent substructures. Then, they generate a functional representation for each drug using those frequent substructures. Further, the functional representation of drug pairs is used to predict DDIs. We reproduce CASTER results for our datasets.
\\

\textit{5. Decagon} Decagon~\cite{Zitnik2018} uses a multi-modal graph consisting of protein-protein interactions and drug-protein targets interactions for DDI prediction. It has an encoder-decoder architecture. The encoder exploits a graph convolution network to generate the representation of drugs by embedding all drug interactions with other entities in it. Then, the decoder takes drug pair representations as input and predicts DDIs with an exact side effect. The same TWOSIDES drug-drug interactions network is used in Decagon. That is why we directly compare our model performances with their reported results for TWOSIDES data instead of reproducing Decagon. However, we do not consider DrugBank data for Decagon as we do not have the additional information (e.g., side effects and target protein) in our DrugBank dataset to construct the multi-modal graph.

\begin{table*}
\centering{
\caption{ \centering{Performance comparisons of \texttt{HyGNN} with baseline models on TWOSIDES dataset.} }
\label{tab5}
\normalsize
\begin{tabular}{  |c | c | c c c  |}
\hline
\cellcolor{gray!60} \textbf{Model}
 & \cellcolor{gray!60}\textbf{Method}
 & \cellcolor{gray!60}\textbf{ F1 }
& \cellcolor{gray!60}\textbf{ ROC-AUC } &\cellcolor{gray!60} {\textbf{PR-AUC} } 
  \\ 
\hline\hline
 & DeepWalk  & 80.35 & 80.36 & 85.19 \\
 {RWE on DDI Graph} & Node2vec & 84.50 & 84.52 & 88.33\\
\hline
 & GCN & 85.34 & 85.38 & 88.87\\
 {GNN on DDI graph} & GraphSage & 85.83 & 85.80 & 89.28 \\
  & GAT & 82.67 & 82.68 & 86.86\\
\cline{1-5}
 & GCN & 53.85 & 54.04 & 66.94\\
 {GNN on SSG graph} & GraphSage & 60.19 & 60.18 & 70.34 \\
  & GAT & 54.25 & 54.37 & 66.85\\
\cline{1-5}
 CASTER & - & 82.35 & 90.45 & 90.58 \\ 
\cline{1-5}
Decagon & - & - & 87.20 & 83.20 \\ 
\cline{1-5}
 & ESPF \& MLP  & 88.79 & 96.01 & 96.30\\
 & ESPF \& Dot & 76.79 & 91.12 & 93.37\\
 \texttt{HyGNN} & \textbf{$k$-mer \& MLP} & \textbf {89.21} & \textbf{96.25} & \textbf{96.53}\\
 & $k$-mer \& Dot & 78.55 & 91.80 & 93.88\\
\hline
\end{tabular}}
\end{table*}

\begin{table*}[t]
\centering{
\caption{ \centering{Performance comparisons of \texttt{HyGNN} with baseline models on DrugBank dataset.} }
\label{tab6}
\normalsize
\begin{tabular}{  |c | c | c c c  |}
\hline
\cellcolor{gray!60}\textbf{Model}
 & \cellcolor{gray!60} \textbf{Method}
 & \cellcolor{gray!60} \textbf{ F1 }
& \cellcolor{gray!60} \textbf{ ROC-AUC } & \cellcolor{gray!60} {\textbf{PR-AUC }} 
  \\ 
\hline\hline
& DeepWalk  & 73.34 & 73.35 & 80.05 \\
{RWE on DDI Graph} & {Node2vec} & 79.52 & 79.54 & 84.56\\
\cline{1-5}
&  GCN & 77.05 & 77.06 & 82.78\\
{GNN on DDI graph} & {GraphSage} & 80.83 & 80.88 & 85.51 \\
&  GAT & 63.84 & 69.75 & 78.52\\

\cline{1-5}
& GCN & 58.00 & 58.04 & 69.11\\
{GNN on SSG graph} & {GraphSage} & 61.10 & 61.15 & 70.64 \\
& GAT & 58.20 & 58.24 & 69.25\\
\cline{1-5}
CASTER & - & 87.36 & 94.27 & 94.20
 \\ 
\cline{1-5}
& ESPF \& MLP  & 92.42 & 97.63 & 97.53\\
& ESPF \& Dot & 83.94 & 95.80 & 96.57\\
\texttt{HyGNN} & \textbf{$k$-mer \& MLP} & \textbf{94.61} & \textbf{98.69} & \textbf{98.68}\\
& {$k$-mer \& Dot} & 87.38 & 97.99 & 98.28\\
\hline
\end{tabular}}
\end{table*}

\begin{figure*}
    \centering{ 
    \includegraphics[width= \textwidth, height= 10cm]{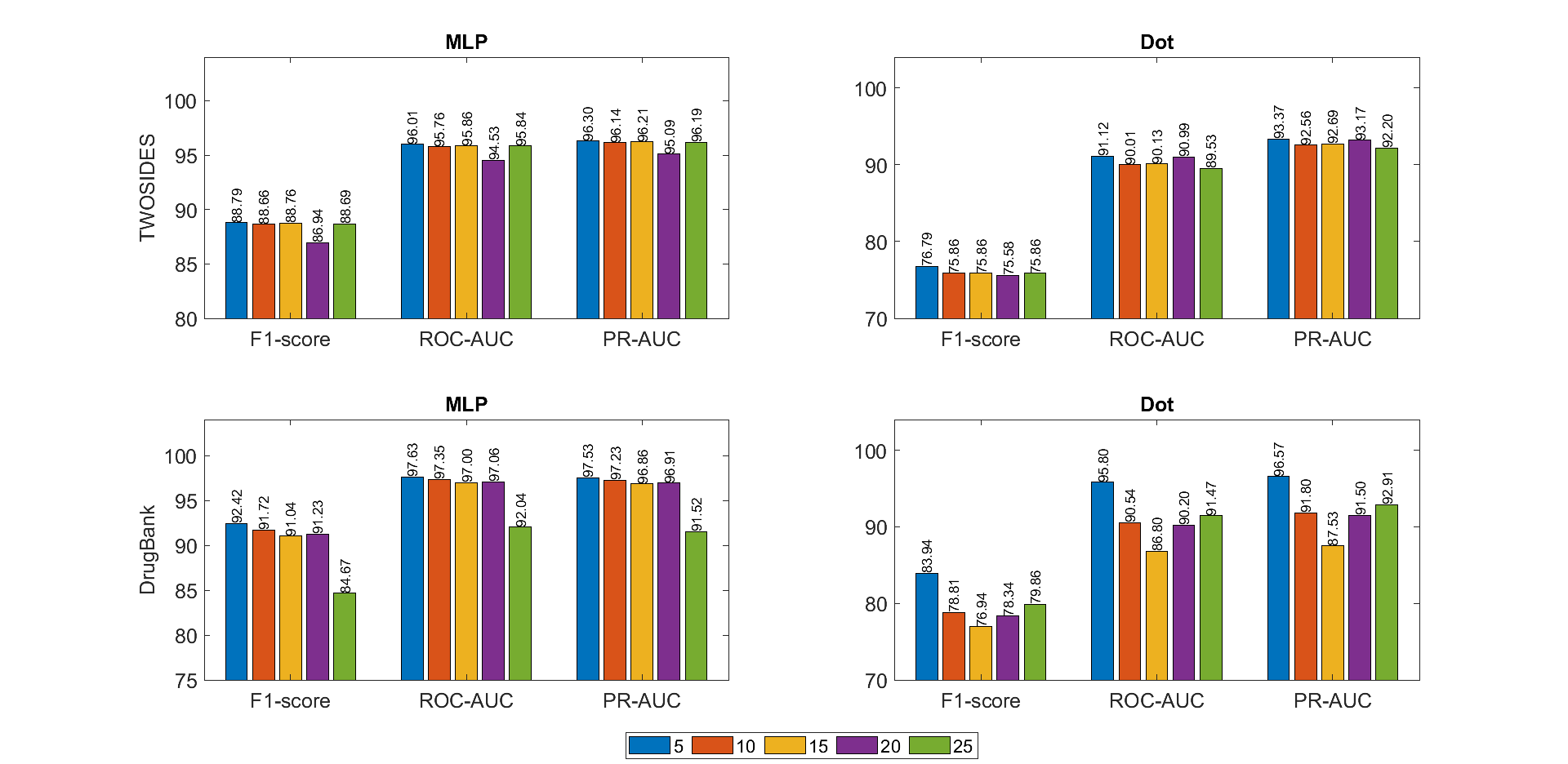}
   \caption{Performance comparison of  models for different frequency thresholds of ESPF.}
    \label{fig:result_graph1}}
    \end{figure*}
    \begin{figure*}
    \centering{ 
    \includegraphics[width=\textwidth, height=10cm]{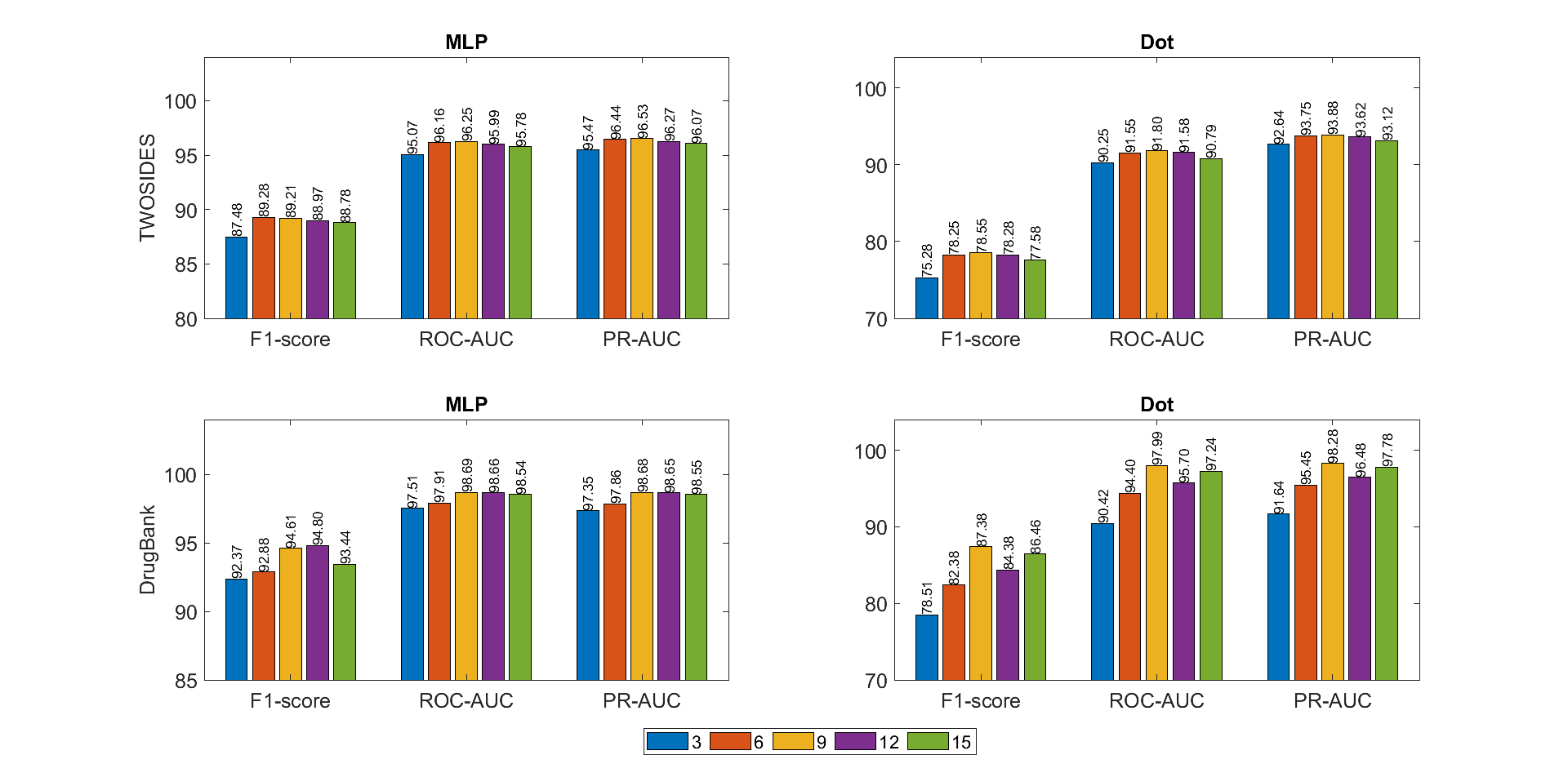}
   \caption{Performance comparison of  models for different sizes of $k$-mer.}
    \label{fig:result_graph2}}
\end{figure*}

\subsection{Results}

\subsubsection{Model Performances} We conduct detailed experiments on our proposed models for two different datasets with different threshold values of $k$-mer and ESPF. Both MLP and dot predictor-based decoder functions are employed individually for each setup to compare their performances. The overall performances are illustrated in Fig. \ref{fig:result_graph1} and Fig. \ref{fig:result_graph2}. Fig. \ref{fig:result_graph1} depicts the model's performance for different ESPF frequency thresholds ranging from 5 to 25. This figure shows that
it has a more significant impact on the TWOSIDES dataset, especially for the Dot decoder function than DrugBank. On TWOSIDES with MLP, it gives similar results till 25, and then it has a considerable decrease for 25. Since we get a significantly less number of substructures, it
would not be enough to learn with those. For DrugBank, it gives similar results for different thresholds of ESPF. In general, frequency threshold 5 gives the best performance for TWOSIDES and DrugBank with MLP and DOT. As with the increment of the frequency threshold, the number of substructures (i.e., nodes) decreases, which could be a potential reason for performance degradation. The best performance for each dataset and
decoder is recorded for a threshold value
of 5.

\begin{figure*}[t]
    \centering
    \includegraphics[width=0.8\textwidth]{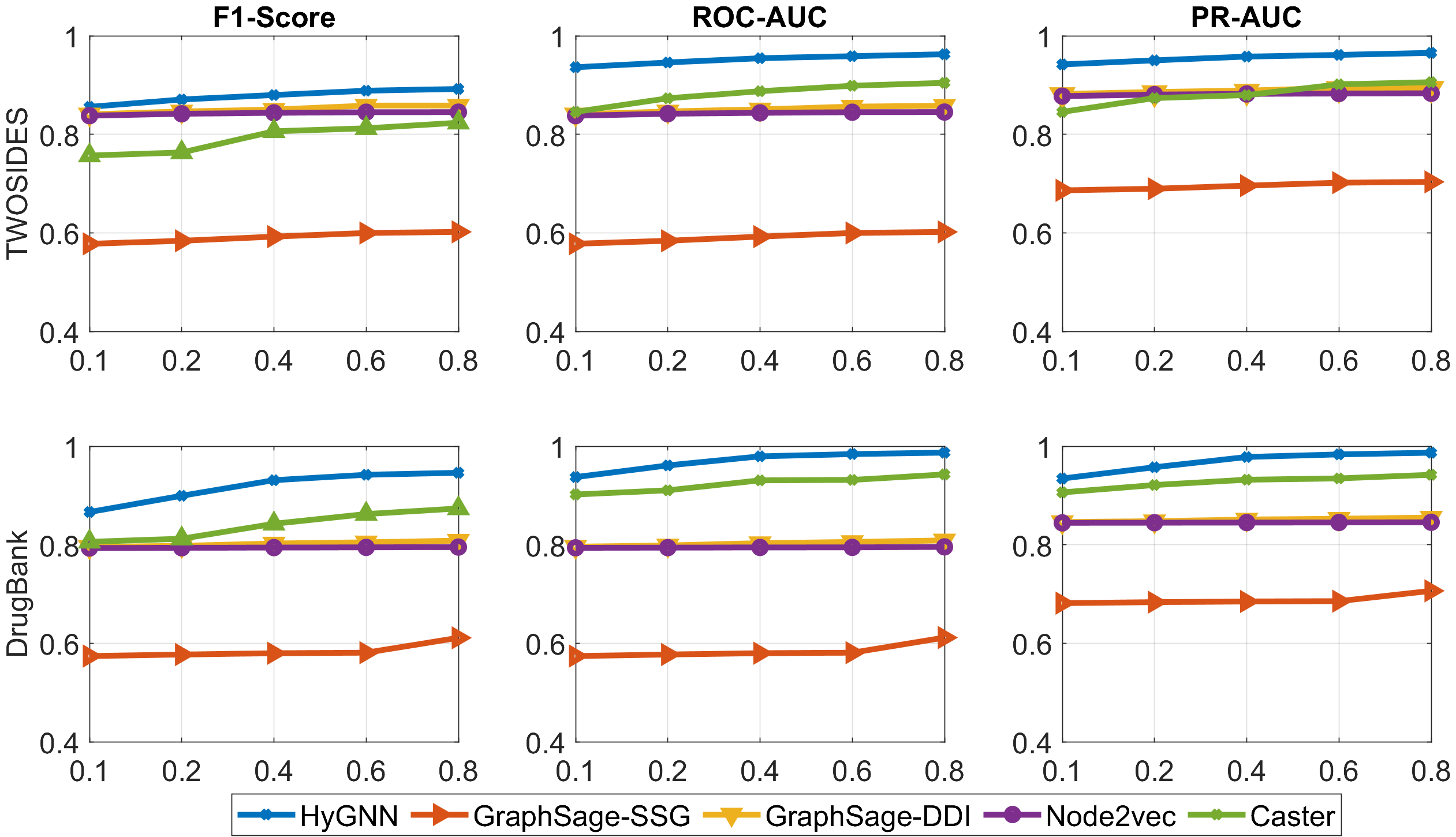}
   \caption{Performance comparison of models for different training sizes where $x$-axes represent the training percentages.}
    \label{fig:result_graph}
\end{figure*}

Fig.~\ref{fig:result_graph2} presents the models' performances for five different $k$ values of $k$-mer ranges from 3 to 15.  Similar to ESPF, the effect of the parameter on the results is higher for TWOSIDES than DrugBank. The reason for this could be that it is smaller than DrugBank, so the graph's size is affecting its results. However, DrugBank is a large dataset with enough training data to get good results, even with a small graph. Here, we can see that with the increment of the size of $k$-mer, the performance of the model increases, especially for TWOSIDES. As with increasing $k$, the number of substructures (i.e., nodes) increases which could be the reason for overall performance improvement. After some point, we will get too many substructures that could put noise into the data and decrease the model's performance. The best performance for each
dataset and decoder are reported with $k=9$ for $k$-mer.

\subsubsection{Comparison with Baselines}
We evaluate our models by comparing their performances with several baseline models and present the results in Table~\ref{tab5} for the TWOSIDES, and Table~\ref{tab6} for the DrugBank dataset. 
As we see in these tables, our models comprehensively outperform all the baseline models. More precisely, in Table \ref{tab5} for the TWOSIDES dataset, \texttt{HyGNN} achieves at least 7\% on F1, 6\% on ROC-AUC, and PR-AUC better performance than other baseline models. While the best model among all the baselines, CASTER, achieves an F1 score of 82.35\%, our \texttt{HyGNN} with $k$-mer \& MLP scores 89.21\% with almost 7\% gain. A similar situation also happens for the other two accuracy measures: ROC-AUC and PR-AUC. 

In Table \ref{tab5}, for the DDI graph, from the GNN models, GraphSage gives the best results with 85.83\%, 85.80\%, and 89.29\% on F1, ROC-AUC, and PR-AUC, respectively. Also, from random walk-based embedding models, Node2Vec gives the best results, which is very similar to GraphSage results with 84.50\% on F1, 84.52\% on ROC-AUC, and 88.33\% PR-AUC scores. For the SSG graph, again, GraphSage gives the best result. In our result comparison table, CASTER is the best competitor of \texttt{HyGNN}. Out of all baseline models, CASTER shows the best performance with ROC-AUC and PR-AUC of above 90\%. Decagon is a multi-modal graph that also exhibits better performance than GNN on SSG.

Table~\ref{tab6} presents the performance comparison of \texttt{HyGNN} with baselines for the DrugBank dataset. As for TWOSIDES, here again, out of three different GNN models on DDI and SSG graphs, GraphSage yields the best result. Similarly, Node2Vec performs better than DeepWalk. CASTER is still the best performer among all baselines, with 87.36\%, 94.27\%, and 94.20\% on F1, ROC-AUC, and PR-AUC, respectively. However, our \texttt{HyGNN} with $k$-mer \& MLP significantly surpasses CASTER with 94.61\% on F1, 98.69\% on ROC-AUC, and 98.68\% on PR-AUC.
As Decagon depends on the drug and other drug-centric information, we could not experiment with it on the DrugBank dataset. 

In summary, \texttt{HyGNN} with $k$-mer gives better results than ESPF. The reason for this could be that with the ESPF, we eliminate many substructures but keep just frequent ones. This may result in losing important ones that are not frequent. However, with $k$-mer, we get all and let the attention models in \texttt{HyGNN} learn which substructures are more important for DDI. 

Moreover, we take the best-performing method from each baseline model, namely Node2Vec from random walk-based embedding, GraphSage from GNN on DDI, GraphSage from GNN on SSG, CASTER, and $k$-mer \& MLP from our \texttt{HyGNN} models. Then, we compare their performances by changing the training sizes from 10\% to 80\% for both datasets. A comparison of performance is outlined in Fig \ref{fig:result_graph}. Results indicate \texttt{HyGNN} to be the best-performing model, and it still gives very good results with small training data. However, decreasing the training size affects the baseline models significantly, especially GraphSage on the SSG model. It is worthy of mention that based on our results, all graph-based models, especially different variants of GNNs, including \texttt{HyGNN} and baselines, have performed fairly well on our data.

\begin{table*}[t]
\centering{
\caption{ \centering{Novel DDI Predictions by \texttt{HyGNN} on TWOSIDES Dataset } }
\label{tab7}
 \normalsize
\begin{tabular}{ | c | c | c | c | c | }
\hline
\cellcolor{gray!60}\textbf{Drug1}
 & \cellcolor{gray!60}\textbf{ Drug2 }
  & \cellcolor{gray!60}\textbf{ TWOSIDES Label  }  & \cellcolor{gray!60}\textbf{ Predicted DDI Score } & \cellcolor{gray!60}\textbf{ DrugBank Label } \\ 
\hline\hline
   Desvenlafaxine & Paroxetine & 0 & 0.9989 & 1\\
  Probenecid &	Metformin &	0 & 0.9931 & 1 \\
  Fluvastatin  & Metronidazole  & 0 & 0.9212 & 1 \\
  Loratadine &	Isradipine & 0 & 0.9703 & 1\\
  Glyburide & Bosentan & 0	& 0.9068 & 1 \\
  Salmeterol  & Dicycloverine & 0 & 0.9189 & 1\\
  Valdecoxib  & Sodium sulfate & 0	& 0.9105 & 1 \\
 Lisinopril  & Naratriptan & 0 & 0.9336 & 1\\
  Bexarotene & Maprotiline & 0 & 9.9993e-10 & 0\\
  Amoxapine & Econazole & 0 & 6.8256e-09 & 0 \\
  Nabilone & Oxaprozin& 0 & 4.1440e-08 & 0\\
 Dexmedetomidine & Carbachol  & 0	& 1.2417e-08 & 0\\

\hline
 
\end{tabular}}
\end{table*}

\begin{table*}[t]
\centering{
\caption{ \centering{Novel DDI Predictions by \texttt{HyGNN} on DrugBank Dataset } }
\label{tab8}
 \normalsize
\begin{tabular}{ | c | c | c | c | c | c | }
\hline

  \cellcolor{gray!60}\textbf{Drug1}
 & \cellcolor{gray!60}\textbf{ Drug2 }
  & \cellcolor{gray!60}\textbf{ DrugBank Label  }  & \cellcolor{gray!60}\textbf{ Predicted DDI Score } & \cellcolor{gray!60}\textbf{ TWOSIDES Label } 
  \\ 
\hline\hline
   Hydroxychloroquine & Loratadine  & 0 & 0.9879 & 1\\
  Dextromethorphan  & Ofloxacin  & 0 & 0.9772 & 1 \\
  Midazolam & Warfarin & 0 & 0.9884 & 1 \\
  Benzthiazide	& Fentanyl & 0 & 5.6989e-14 & 0\\
  Labetalol & Levonorgestrel &	0 & 9.1049e-07 & 0 \\
 Cefprozil & Disulfiram & 0		& 1.0882e-11 & 0\\
\hline
 
\end{tabular}}
\end{table*}

\begin{table*}[t]
\centering{
\caption{ \centering{Performance of \texttt{HyGNN} for New Drugs} }
\label{tab9}
\normalsize
\begin{tabular}{  |c | c | c c c  |}
\hline
\cellcolor{gray!60}\textbf{Dataset}
 & \cellcolor{gray!60} \textbf{Unseen Node}
 & \cellcolor{gray!60} \textbf{ F1 }
& \cellcolor{gray!60} \textbf{ ROC-AUC } & \cellcolor{gray!60} {\textbf{PR-AUC }} 
  \\ 
\hline\hline
{TWOSIDES} & {5\%} & 72.75 & 78.25 & 85.64\\
\cline{1-5}
{DrugBank} & {5\%} & 65.23 & 70.84 & 78.04 \\
\hline
\end{tabular}}
\end{table*}

Hypergraphs are used in a wide range of scientific fields. Hypergraphs are a natural method to illustrate shared group relationships. Through a hypergraph structure, \texttt{HyGNN} is able to capture higher-order correlations between data (i.e., triadic, tetradic, etc.). Furthermore, employing an attention mechanism makes it more robust by giving more weight to important substructures while learning representations of drugs. Though GAT has attention architecture as well, it can not discover the important edges. The main strength of our \texttt{HyGNN} is the proposed \textit{hypergraph edge encoder} that has two levels of attention mechanism. At first, it aggregates the hyperedges to generate the representation of the node. While aggregating, it imposes more attention on the important hyperedges. Similarly, to generate the representation of a hyperedge, it aggregates the nodes' information with much attention to the important ones. 

Moreover, \texttt{HyGNN} has a decoder function, and we learn all the parameters of the encoder and decoder simultaneously during training. From Table \ref{tab5} and Table \ref{tab6}, we can see \texttt{HyGNN} with $k$-mer \& MLP performs better than dot product. $k$-mers are $k$-length substrings included inside a biological sequence. A bigger $k$-mer is preferable since it ensures greater uniqueness in the base sequences that will create the string. Larger $k$-mer sizes aid in the elimination of repetitive substrings. Moreover, MLP Predictors are well-suited for classification problems in which data is labeled. They are extremely adaptable and may be used to learn a mapping from inputs to outputs in general. Additionally, it generates superior results compared to dot predictor since it has trainable parameters that are learned throughout the training.

\subsubsection{Case Study - Prediction and Validation of Novel DDIs} We evaluate the effectiveness of \texttt{HyGNN} model on novel DDIs prediction. We select some drug pairs from TWOSIDES. None of these drug pairs have  DDI info in TWOSIDES but have DDI info in DrugBank. Then we train our \texttt{HyGNN} using TWOSIDES and make predictions for those pairs. Following that, we collect predicted scores for those drug pairs as shown in Table \ref{tab7}. From this table, we see that for the first eight drug pairs, though the TWOSIDES label for each of these pairs is zero, we get predicted scores above 90\%  for each pair, which shows there is a high chance that each pair will interact between them. To further validate it, we cross-check our predicted score with DrugBank, which says each of these eight drug pairs interacts between them. Moreover, for the last four-drug pairs of Table \ref{tab7}
the predicted scores are minimal, and TWOSIDES, and DrugBank both say they don’t interact. Similarly, six drug pairs are selected from DrugBank having no DDI info in DrugBank but in TWOSIDES as shown in Table \ref{tab8}, then \texttt{HyGNN} is trained using DrugBank data and validated the predicted scores by TWOSIDES.

\subsubsection{Case Study- DDI Prediction for New Drugs} 

Making DDI predictions for new drugs could be more challenging than existing drugs. Since the model does not learn based on the SMILES strings of new drugs.
To show the effectiveness of our model for new drugs, at first, we randomly select a 5\% drug from a dataset and completely remove these drugs' information from the corresponding train set and keep those drugs' information only in the test set. These selected 5\% drugs can be considered new drugs. The experimental results for both datasets with new drugs are shown in Table \ref{tab9}. As we see in the table, our model predicts DDIs effectively for both datasets.

\section{Conclusion}
\label{sec:con}
In this paper, we propose a novel GNN-based framework for DDI prediction based on the chemical structures (SMILES) of drugs. In contrast to existing graph-based models, we utilize a novel hypergraph structure to depict higher-order structural similarities between drugs. Following that, we propose a Hypergraph GNN model with an encoder-decoder architecture to learn the drug representation for DDI prediction. We develop a hypergraph edge encoder to construct drug embeddings and a decoder with drug representations to predict a score for each drug pair, indicating whether two drugs interact. Finally, with several experiments, we show that our method outperforms different types of baseline models and also it is able to predict DDIs for new drugs. 

As future work, we plan to extend our model to address other problems in bioinformatics, like protein-protein interaction prediction, drug repurposing, and sequence classification. 

\section*{Acknowledgment}
This work is funded partially by National Science Foundation under Grant No 2104720. 

\bibliography{main}
\end{document}